\documentclass[twocolappendix]{emulateapj}
\slugcomment{}

\usepackage{amsmath}
\usepackage{natbib}
\usepackage[para]{threeparttable}

\begin{document}

\shortauthors{Shah \& Nelemans}
\shorttitle{Astrophysics of Galactic binaries}

\title{Constraining parameters of white-dwarf binaries using
  gravitational-wave and electromagnetic observations}

\author{Sweta Shah$^{1,2}$ and Gijs Nelemans$^{1,2,3}$}

\affil{ $^1$Department of Astrophysics/ IMAPP, Radboud University Nijmegen,
  P.O. Box 9010, 6500 GL Nijmegen, The Netherlands\\
	$^2$Nikhef – National Institute for Subatomic Physics,  Science
  Park 105,  1098 XG Amsterdam, The Netherlands \\
	$^3$Institute for Astronomy, KU Leuven, Celestijnenlaan 200D, 3001
  Leuven, Belgium\\	}

\email{s.shah@astro.ru.nl}
\date{\today}

\begin{abstract} 
The space-based gravitational wave (GW) detector,
  \emph{evolved Laser Interferometer Space Antenna} (eLISA) is
  expected to observe millions of compact Galactic binaries that
  populate our Milky Way. GW measurements obtained from the eLISA
  detector are in many cases complimentary to possible
  electro-magnetic (EM) data. In our previous papers, we have shown
  that the EM data can significantly enhance our knowledge of the
  astrophysically relevant GW parameters of the Galactic binaries,
  such as the amplitude and inclination. This is possible due to the
  presence of some strong correlations between GW parameters that are
  measurable by both EM and GW observations, for example the
  inclination and sky position. In this paper, we quantify the
  constraints in the physical parameters of the white-dwarf binaries,
  i.e. the individual masses, chirp mass and the distance to the
  source that can be obtained by combining the full set of EM
  measurements such as the inclination, radial velocities, distances
  and/or individual masses with the GW measurements. We find the
  following $2-\sigma$ fractional uncertainties in the parameters of interest.
  The EM observations of distance constrains the the chirp mass to
  $\sim 15-25 \%$, whereas EM data of a single-lined spectroscopic
  binary constrains the secondary mass and the distance with factors
  of 2 to $\sim 40 \%$. The single-line spectroscopic data
  complemented with distance constrains the secondary mass to $\sim
  25-30\%$. Finally EM data on double-lined spectroscopic binary
  constrains the distance to $\sim 30\%$. All of
  these constraints depend on the inclination and the signal strength of the 
  binary systems. We also find that the EM information on distance and/or the
  radial velocity are the most useful in improving the estimate of the
  secondary mass,
  inclination and/or distance.
\end{abstract}
\maketitle
\section{Introduction}
Gravitational wave (GW) observations and electro-magnetic (EM)
observations can be used to study compact Galactic binaries
independently and often these two ways provide different measurements
of the same system. There are about $\sim$50 of these binaries that
have been studied in the optical, UV, and X-ray wavelengths 
\citep[e.g.][]{2010ApJ...711L.138R}. This number is expected to grow
by a factor of $\sim$100 \citep{2012ApJ...758..131N} 
when a space-based gravitational wave (GW) observatory like the
recently eLISA\footnote{In preparation by ESA, expected
  launch in $\sim$2034} will be in operation. This detector is expected
to observe millions of compact Galactic binaries with periods shorter
than about a few hours \citep{2009CQGra..26i4030N,
  2013GWN.....6....4A}, amongst other astrophysical sources. Of those
millions of binaries we will be able to resolve several thousands. 
It has been shown \citep[Paper I, hereafter]{2012A&A...544A.153S} 
that for a non-eclipsing binary system (for example AM~CVn), its EM
measurement of the inclination, $\iota$ can improve the error on the GW
amplitude ($\mathcal{A}$) significantly depending on the strength of the GW
signal and the magnitude of the EM uncertainty in the inclination. $\mathcal{A}$ is
a GW parameter which is given by a combination of the 
masses, orbital period and distance to the source:
\begin{equation}
\mathcal{A} = \frac{4 (G \mathcal{M}_c)^{5/3}}{c^4 d}(\pi f)^{2/3},
\label{eq:amp}
\end{equation}
where, $d$ is the distance to the source, $f$ is the source's GW
frequency ($2/P_{\mathrm{orb}}$), and $\mathcal{M}_c$ is the chirp mass defined as:  
\begin{equation}
\mathcal{M}_c \equiv (m_1\:m_2)^{3/5}/(m_1+m_2)^{1/5}.
\label{eq:chirp_mass}
\end{equation}
From the GW observations alone, one typically cannot
measure the individual masses or the distance since they are
degenerate via Eqs. \ref{eq:amp}, \ref{eq:chirp_mass}. In the rare
cases that a precise
orbital decay ($\dot{f}$), can be measured from GW data then the
distance can be estimated (with the assumption that the frequency
evolution is dictated by GW radiation only) 
by determining $\mathcal{M}_c$ from the measured $f$ and $\dot{f}$ via
the equation \citep{1963PhRv..131..435P}:
\begin{equation}
\dot{f} = \frac{96 \: \pi}{5}\frac{G^{5/3}}{c^5}\:(\pi\;\mathcal{M}_c)^{5/3}\:{f}^{11/3}.
\label{eq:fdot}
\end{equation}
For the compact binaries that have been observed with the optical
telescopes, a subset of which will also be detected by eLISA, their EM
data often provide measurements of the orbital period $P_{\mathrm{orb}}$,
the primary mass ($m_1$), sometimes the secondary mass 
($m_2$), the distance ($d$) and the radial velocity
amplitude ($K_i$). We use these measurements for a few binaries to show the
quantitative improvements in their GW and other physical parameters. Many of these
binaries can/could still be found 
electromagnetically before or after eLISA discoveries.

We have previously shown that knowing sky positions from EM data can
improve the GW uncertainties on $\mathcal{A}$ and $\iota$ depending
on the particular geometry and orientation of the binary systems
\citep{2013A&A...553A..82S}. Thus, so far we have quantitatively studied the
improvement factors in the uncertainties of the parameters that can be
gained from \emph{prior} knowledge of parameters which are common to
both GW and EM observations, for example inclination, and sky
position. 

In this paper we go beyond constraining \emph{only} those GW
parameters which are \emph{also} measured independently from the EM
data. We explore various combinations of \emph{any} possible EM 
observations and the GW measurements in constraining the
\emph{useful parameters} of the binaries that are astro-physically
interesting, for example the individual masses. Because their GW
signal is significantly affected we consider high-inclination
(sometimes eclipsing) and (low inclination) binary systems. We review
the GW data analysis methods in Sect. 2. In 
Sect. 3, we explore the information gained by combining EM
measurements in different ways where the EM data can be the radial
velocity of one of the binary components, $K_i$, $m_1, \: m_2$, $d$,
and $P_{\mathrm{orb}}$. Specifically, we classify various combinations
into a number of scenarios in discussing the parameter constraints. 
\begin{table*}[!t]
\centering
\begin{threeparttable}
\caption{GW parameter values of J0651}
\label{tab:verf_bin}
\begin{tabular}{c c c c c c c c c c c c }
\hline \hline
&$\mathcal{A}\;\;[\times 10^{-22}]$ &$\mathcal{M}_c \;\;[M_{\odot}]$ &$\phi_0$[rad] & $\cos\iota$ & $f \;\;[\times10^{-3}]$[Hz] &$\psi$[rad] & $\sin\beta$ & $\lambda$[rad] & $\mathrm{S/N} $ \\
\hline
J0651 &$1.67$\tnote{a}, $\; 6.71$\tnote{b}  &$0.32$\tnote{a}, $\;  0.70$\tnote{b}  & $\pi$ & $0.007$ & $2.61$ & $\pi/2$ & $0.101$
&$1.77 $ & $\sim 13$\tnote{a}, $50$\tnote{b} 
 \\\hline
      \end{tabular}
    \begin{tablenotes}
        \item[a]for $m_1 = 0.25 M_{\odot}$, $m_2 = 0.55 M_{\odot}, d = 1.0$ kpc
        \item[b] for $m_1 = 0.8 M_{\odot}$, $m_2 = 0.8 M_{\odot}, d = 1.0$ kpc 
\end{tablenotes}
\end{threeparttable}
\end{table*}
\section{Parameter uncertainties from GW observations}
For our analyses below, we consider one of the eLISA
\emph{verification binaries} J065133.33+284423.3 (J0651, hereafter;
\cite{2011ApJ...737L..23B}). We also consider a second (hypothetical)
system with higher masses which we will refer to as ``the high mass
binary''. Their GW parameter values are listed in Table 1. Before
looking at the EM data we briefly recap out GW data analysis
method. We have used Fisher matrix studies
\citep[e.g][]{1998PhRvD..57.7089C} to extract the GW parameter 
uncertainties and correlations in the GW parameters that describe the
compact binary sources. Our method and application of Fisher
information matrix (FIM) for eLISA binaries together with their signal
modeling and the noise from the detector and the Galactic foreground 
have been described in detail in Paper I. Most of the binaries will be
monochromatic sources and such sources are completely characterized by
a set of seven parameters, dimensionless amplitude ($\mathcal{A}$),
frequency ($f$), polarization angle ($\psi$), initial GW phase
($\phi_{0}$), inclination ($\cos \iota$), ecliptic latitude ($\sin
\beta$), and ecliptic longitude ($\lambda$). From the GW signal of a
binary and a Gaussian noise we can use FIM to estimate the parameter
uncertainties. The inverse of the FIM
is the variance-covariance matrix whose diagonal elements are the GW
uncertainties and the off-diagonal elements are the correlations
between the two parameters. We do the GW analyses of the above
mentioned binaries for eLISA observations of two years. We
note that Fisher-based method is a quick way of computing  parameter
uncertainties and their correlations in which these uncertainties are
estimated locally at the true parameter values and therefore by
definition the method cannot be used to sample the entire 
posterior distribution of the parameters. Additionally Fisher-based
results hold in the limit of strong signals with a Gaussian noise
\citep[e.g.][see also Appendix]{2008PhRvD..77d2001V}. 

\begin{figure}[!h]
\centering 
\includegraphics[width=\columnwidth]{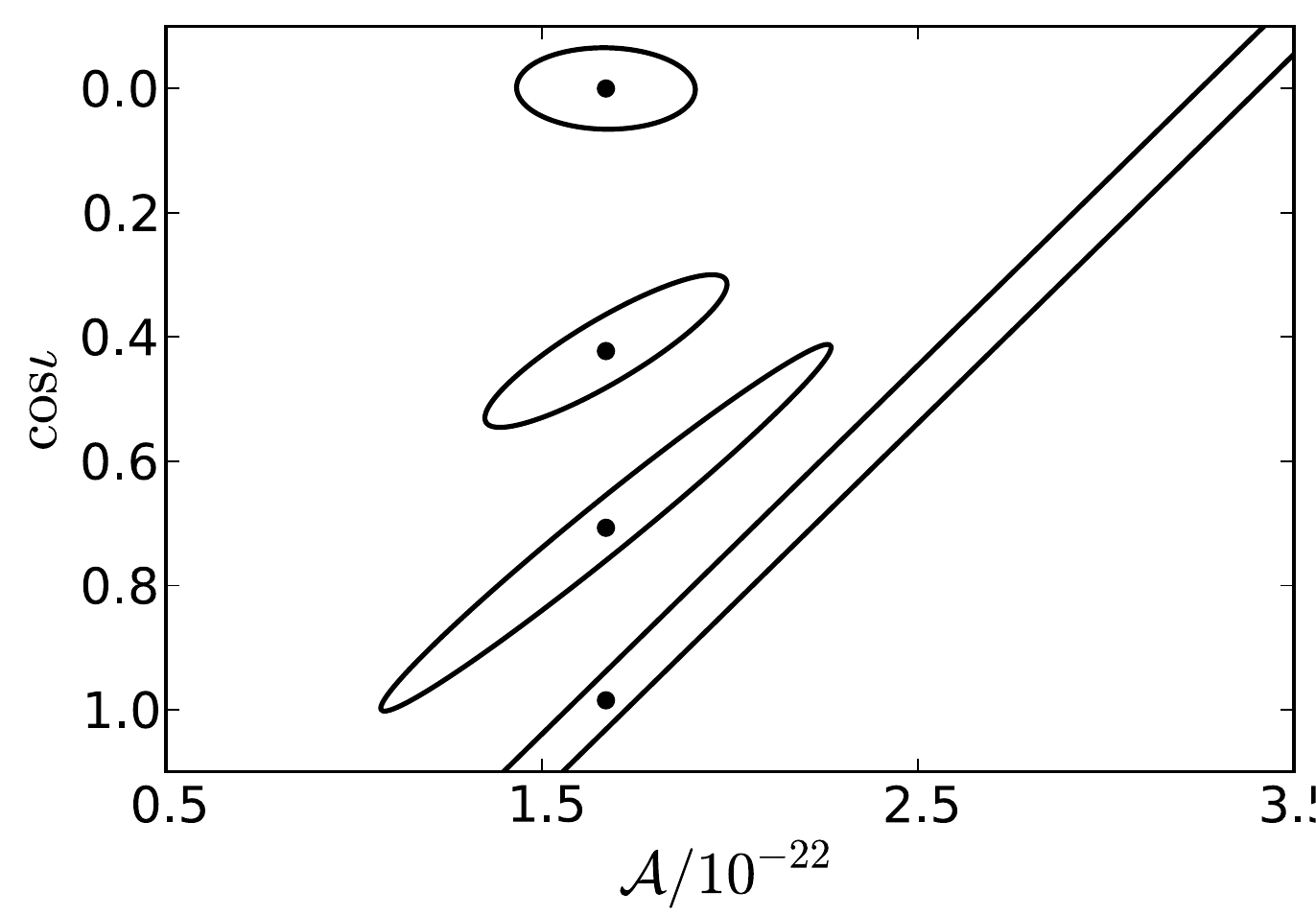}
\caption{Two-dimensional error ellipses of $\mathcal{A}$ and
  $\cos\iota$ extracted from the variance-covariance matrix for J0651
  binary system with varying orientation in its $\iota$. The
  distributions with larger to smaller ellipses correspond to $\iota =
  10^{\circ}, 45^{\circ}, 65^{\circ}, 90^{\circ}$ respectively. The black dots are
  the GW parameter values (see Table 1) about which the FIM is evaluated for the
  corresponding orientations.}
\label{fig:dist}
\end{figure}
\begin{figure}[!h]
\centering 
\includegraphics[width=\columnwidth]{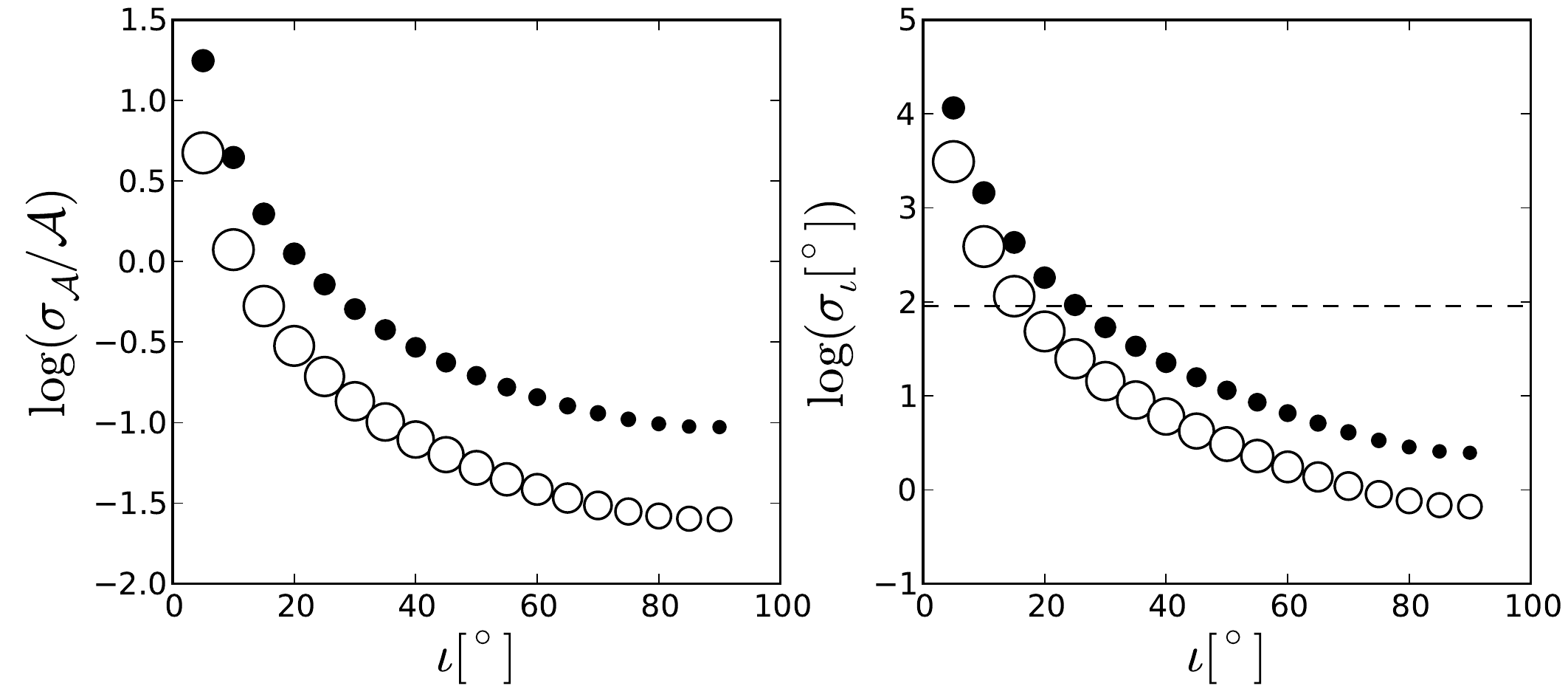}
\caption{GW uncertainties in amplitude and inclination for J0651 as
  a function of inclination calculated from Fisher matrices. Filled circles
  are for J0651 with $m_1, m_2 = 
  0.25, 0.55 M_{\odot}$ and open circles are for the case of the high
  mass binary with $m_1, m_2 =
  0.8, 0.8 M_{\odot}$. The size of the marker represents the S/N at
  each inclination. The dashed line on the right-panel marks the
  unphysical values for the inclination (see text).}
\label{fig:uncertainties}
\end{figure}
The two-dimensional GW distribution in amplitude and inclination given
by the variance-covariance matrix for J0651 parameters are shown in
Figure~\ref{fig:dist} for a number of inclinations. The largest and
most highly correlated distribution is that with $\iota = 10^{\circ}$
and the weakest correlation is that with $\iota = 90^{\circ}$. The
behavior of these distributions reflect the strength of the
correlation between amplitude and inclination. As
discussed in Paper I, the low inclination systems $\iota \leq
45^{\circ} $ have very similar signal shapes, whereas
systems with high inclinations are distinguishable by both the shape
and structure for small differences in inclinations. Thus, for
low-inclination binaries a small change in $\iota$ is
indistinguishable from a small change in its $\mathcal{A}$. On the
other hand for high-inclination binaries a 
small change in inclination produces a noticeably different signal
explaining the uncertainties in $\mathcal{A}$ and $\iota$ becoming
large to small with increasing inclination. The GW
uncertainties for the amplitude and inclination as a function of
inclination are  shown in Figure~\ref{fig:uncertainties} for J0651 (in
filled circles) and the high mass binary (in open circles).
The strong increase in uncertainty trends for low inclination systems
is due to the correlation between amplitude and inclination
\citep{2012A&A...544A.153S}. Clearly the high mass binary has larger
S/N which gives smaller uncertainties in both of its parameters shown
in open circles in the figure compared to that of
J0651. Observe that inclination is a cyclic parameter and is
bounded between $0^{\circ}\leq\iota\leq90^{\circ}$ and yet we get very
large uncertainties from Fisher matrix for lower inclinations systems
shown in the right panel of in Figure~\ref{fig:dist}. This is due to
the fact that Fisher matrix methods are based on the linearised signal
approximation as a result of which it is not sensitive to the bounded
parameters that describe the signal model
\citep{2008PhRvD..77d2001V}. In other words in FIM one 
computes the uncertainties in parameters based on variation of the
signal with respect to the parameters at the true parameter values and
the fact that far away from the true value the parameter has a bound 
is not taken into account by the FIM. When the uncertainty in a bounded
parameter exceeds its physically allowed range, it means the quantity
cannot be determined from GW data analysis. The dashed line in in
Figure~\ref{fig:uncertainties} indicates the value (at $90^{\circ}$)
beyond which the uncertainties in $\iota$ imply unphysical values for 
the inclination. Since the low inclination systems on the left-side of
the plot are affected by this, corrections have to be applied to the
corresponding (over-estimated) uncertainties in amplitude in the left
panel by discarding the unphysical range in the inclination 
\citep{2013A&A...553A..82S}. One way to correct these
unphysical values is by taking a rectangular prior on the
inclination. This in effect will cut off the posterior distribution in
the parameters at the physical bounds described by the prior. Note
that cutting off the error ellipses at lower inclinations in
Figure~\ref{fig:dist} is reasonable because taking strict bounds far
away form the real value about which Fisher uncertainties are computed
will not change the shape and slope significantly. The cut off in the
posterior distributions due to rectangular priors will skew the means
of the paramter distributions away from the real value
\cite[][Eq. C4]{2013PhRvD..88h4013R}.  Furthermore we stress the fact
that the Fisher matrix method is an \textit{estimate} and cannot,
follow the posterior in detail (see Appendix).

The normalized correlations between all
the seven parameters for an eclipsing and non-eclipsing orientations
of J0651 are listed in the variance-covariance matrices (VCM) in the
Appendix. We will make use of these parameter uncertainties and their
corresponding correlations when combining with various EM data in Sect. 3. 
\subsection{GW information only}
\begin{figure}[!h]
\centering
\includegraphics[width=\columnwidth]{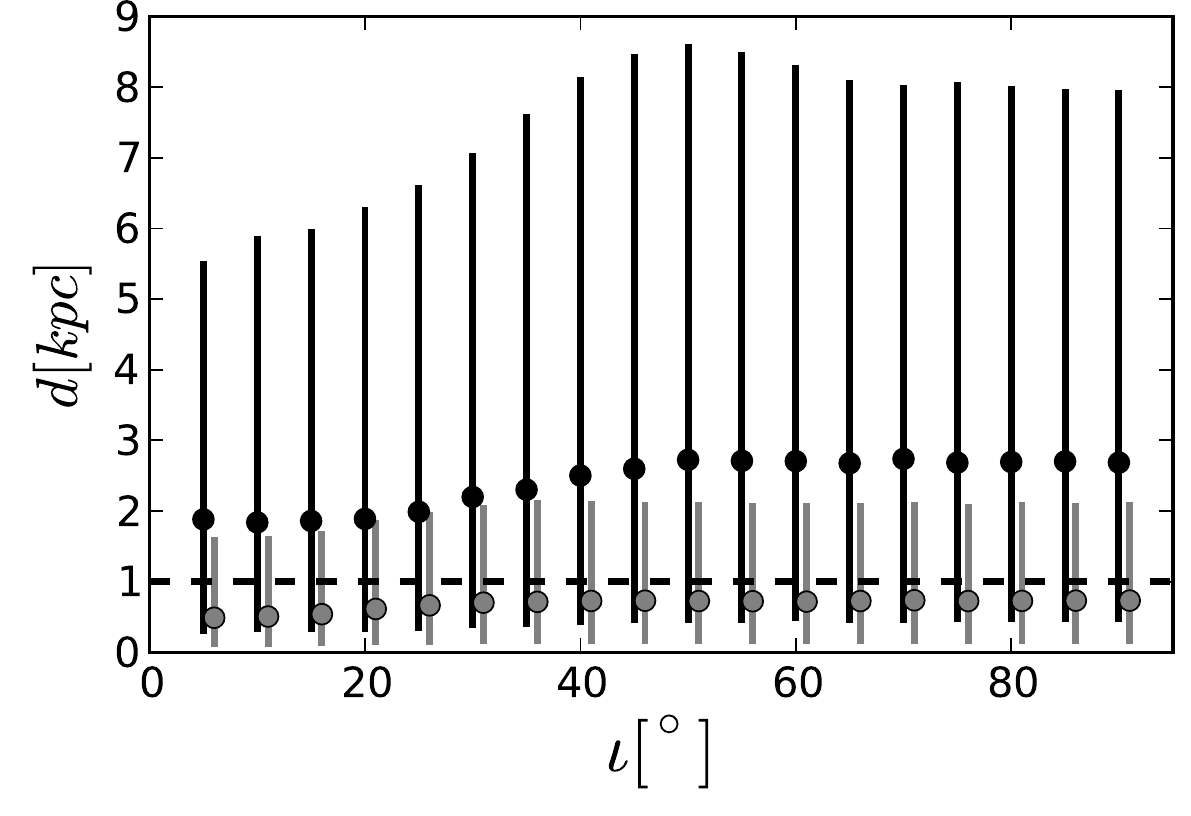}
\caption{\emph{GW data only}: $95$ percentile in distance assuming
  finite chirp mass for J0651 in black lines and the high mass binary
  in grey lines. The dashed line (in black) is the true value. For
  clarity the constraints for the high mass binary are shifted to the
  right. We do this for all the cases below. }
\label{fig:group0}
\end{figure}

\begin{figure*}
\centering 
\includegraphics[width=17cm]{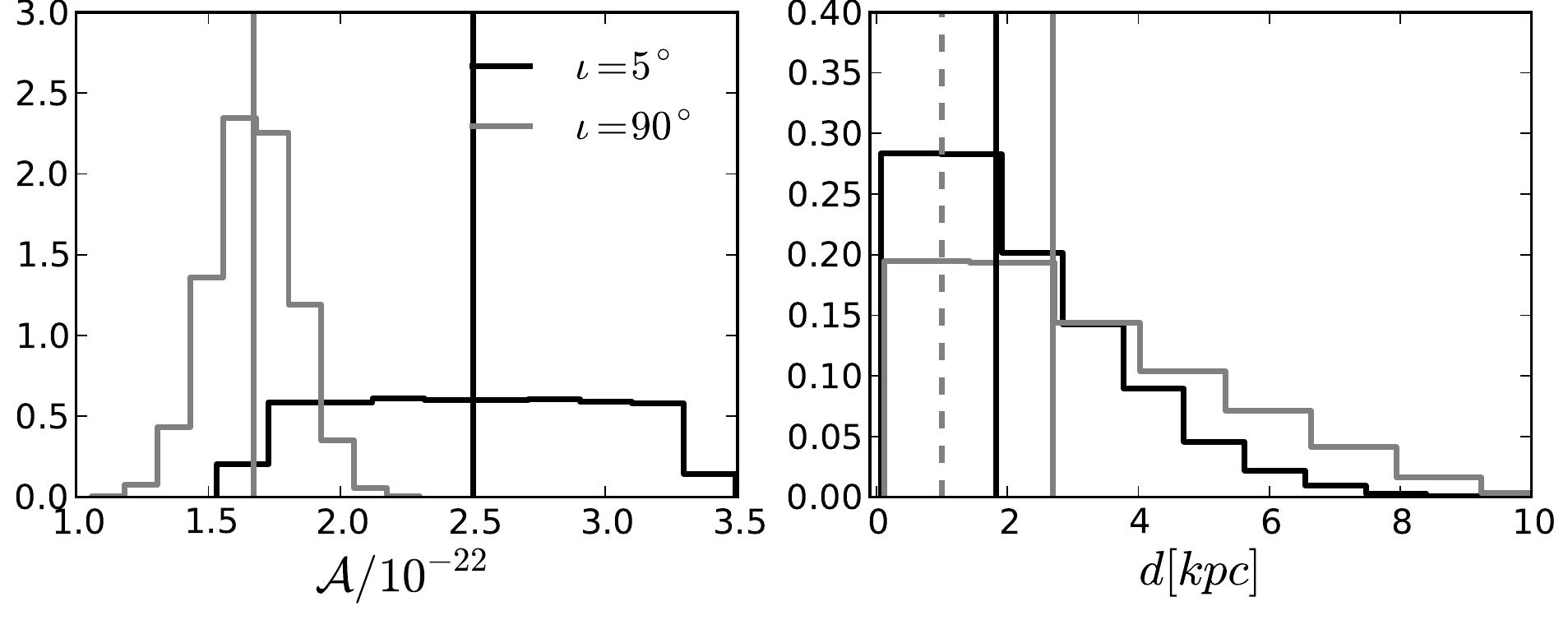}
\caption{\emph{GW data only}: Left: Example of 1D distributions in $\mathcal{A}$ from
  GW data for two inclinations as labeled and their corresponding
  distributions. Right: Assuming a finite $\mathcal{M}_c$, this gives
  corresponding distributions in distance. The solid lines (in grey
  and black) are the medians of the distributions. The real values of
  the parameters are shown in vertical dashed lines (in grey). Note
  that in the left panel the real value is the same as the median of
  the grey distribution and thus they overlap.}
\label{fig:group0_exp}
\end{figure*}

We start by considering the case where we only have the GW data. From
the GW observations, the astrophysical parameters of interest for a 
monochromatic source are its $f$, $\mathcal{A}$ and $\iota$. From the GW
data analysis the frequency of the source will be very well
determined, $\sigma_{f}/f \sim 10^{-6}$ Hz (e.g.\ Paper I) for a $10^{-3}$Hz source, so we
consider that $f$ is essentially known with a fixed value. Given that
most of the binaries that we will observe with eLISA will be binary
white-dwarfs (WD) \citep{2001A&A...375..890N}, we can 
restrict their masses to $m_i \; \epsilon \;[0.1, 1.4] \:
M_{\odot}$. For simplicity we take \emph{uniform}
distributions for both masses in this range. This provides a distribution
in the system's chirp mass, which will provide an upper limit on the
distance for the source. In Figure~\ref{fig:group0} we show these
estimates in $d$ with their $95$ percentile (or $2-\sigma$
uncertainties) as a function of inclination for both J0651 (in black) and
for the high mass binary with equal high-mass components (in grey). 
The dashed line (in black) is the real value of the distance for both systems. The
lower medians of distances at the lower inclinations for both systems
are explained by the fact that at $\iota = 5^{\circ}$, the GW
distribution of $\mathcal{A}$ has a long uniform tail. This is
shown in Figure~\ref{fig:group0_exp} where we compare the
distributions of $\mathcal{A}$ for two inclinations: $5^{\circ}$ (in black),
and $90^{\circ}$ (in grey) in the left panel. For a fixed distribution of
$\mathcal{M}_c$, the corresponding distributions in $d$ are shown in
the right panel where the solid vertical lines are the distribution
medians and dashed vertical line is the real value. We can see
that  $\mathcal{A}^{5^{\circ}}_{\mathrm{median}} >
\mathcal{A}^{90^{\circ}}_{\mathrm{median}}$ thus giving 
$d^{5^{\circ}}_{\mathrm{median}} < d^{90^{\circ}}_{\mathrm{median}}$
\begin{figure*}
\centering 
\includegraphics[width=17cm]{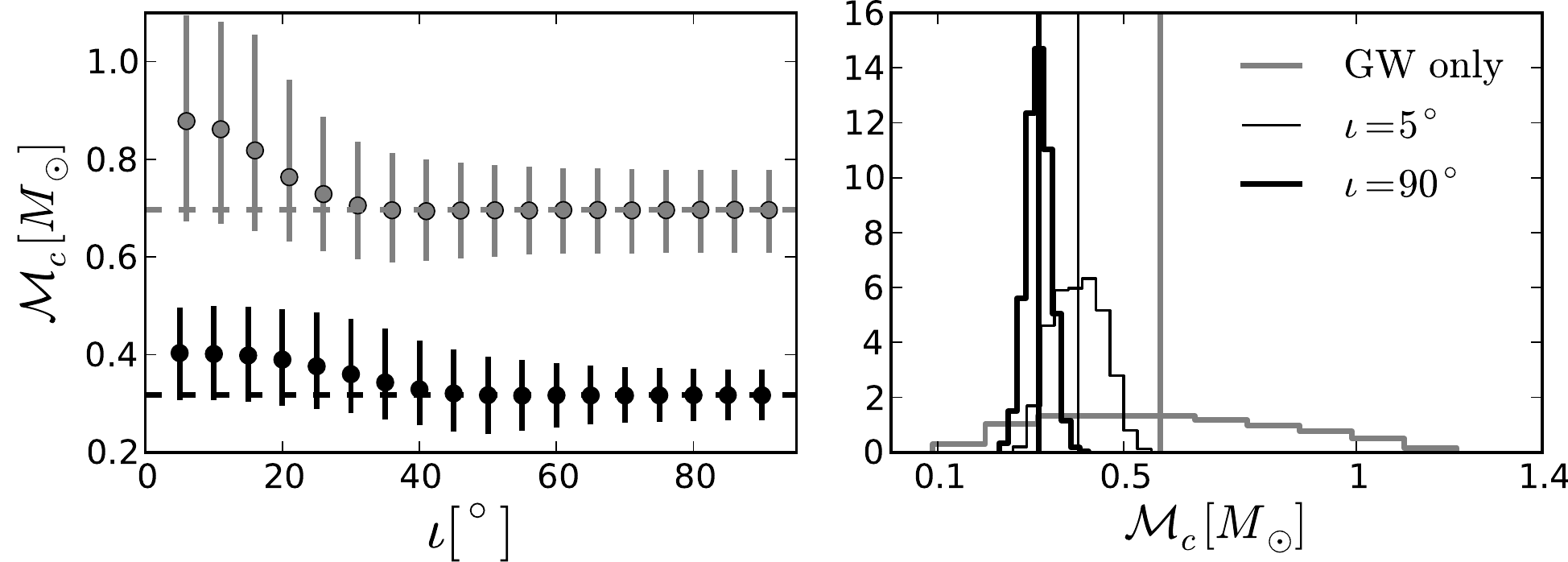}
\caption{\emph{Scenario 1 (known distance)}: Left: $95$ percentile in chirp mass given
  GW data on $\mathcal{A}$ and EM data on distance for J0651 (in black) and high
  mass binary (in grey). Right: Example of
  1D distributions in the $\mathcal{M}_c$ for two inclinations as
  labeled for the J0651. The solid lines are the medians of the distributions
  whereas the dashed line (over-plotted on thick black solid line) is the
  real value of the $\mathcal{M}_c$. For comparison, the
  $\mathcal{M}_c$ computed from uniform distribution of masses is shown in grey.}
\label{fig:group1}
\end{figure*}
via Eq.\,\ref{eq:amp} for a fixed $\mathcal{M}_c$. Also, observe that
the median distances are over-estimated for J0651 for
all inclinations and this is because the real value of the median in
$\mathcal{M}_c$ is much smaller than that computed from uniform
distributions in $m_i$, which is the same for 
all inclinations. Whereas for the high mass binary
the computed median in $\mathcal{M}_c$ is close to its real value thus
translating into smaller offsets in the median distances in
Figure~\ref{fig:group0}. In the figure, the $95$ percentiles in the
distance slightly increase as a function of inclination even though  
the uncertainties in $\mathcal{A}$ has the opposite behavior (see
Figure~\ref{fig:uncertainties}). This is because the $\mathcal{M}_c$
has a very large fractional uncertainty compared to that of the 
$\mathcal{A}$ and thus the relative error uncertainties in the chirp
mass dominates those in the distance, which remain roughly constant
for all inclinations. 
\section{Combining EM \& GW observations}
In all the various scenarios we analyze below, we take the EM 
parameters with an uncertainty of $10\%$ which is inspired by
observational uncertainties of J0651. This binary is a well known
EM source and also a guaranteed source for eLISA. J0651 is an eclipsing
system and such an orientation of a nearby binary allows accurate EM
measurements of it's orbital parameters, and the masses (accuracies of
$\sim 15\%$ (primary mass); $8\%$ (secondary mass)) from observing the
spectra, radial velocities and eclipses
of each star by the other \citep{2011ApJ...737L..23B}. Furthermore its rate of
change of orbital period has also been measured from follow-up high speed 
photometry from $\sim 1$ yr. worth of data  to an accuracy of $\sim
30\%$ \citep{2012ApJ...757L..21H}, and this will improve in the course
of time.  
In this section we classify specific (possible) scenarios where we could
have one or more EM data on the white dwarf binary parameters. We
explicitly state how much the knowledge of any of the various
parameters that describe the \emph{physical} properties of a binary
system can be further improved if we can fold in various combinations
of the existing EM and/or GW observations. We construct three specific
scenarios below based on the typical knowledge from the EM
observations:
\begin{itemize}
\item EM data on distance
\item Single-line spectroscopic data (complemented with or without the
  distance measurement)
\item Double-line spectroscopic data
\end{itemize}
In all the scenarios the GW information on 
amplitude, inclination and frequency from Sect. 2.1 are used. 
\begin{figure*}[t]
\centering 
\includegraphics[width=17cm]{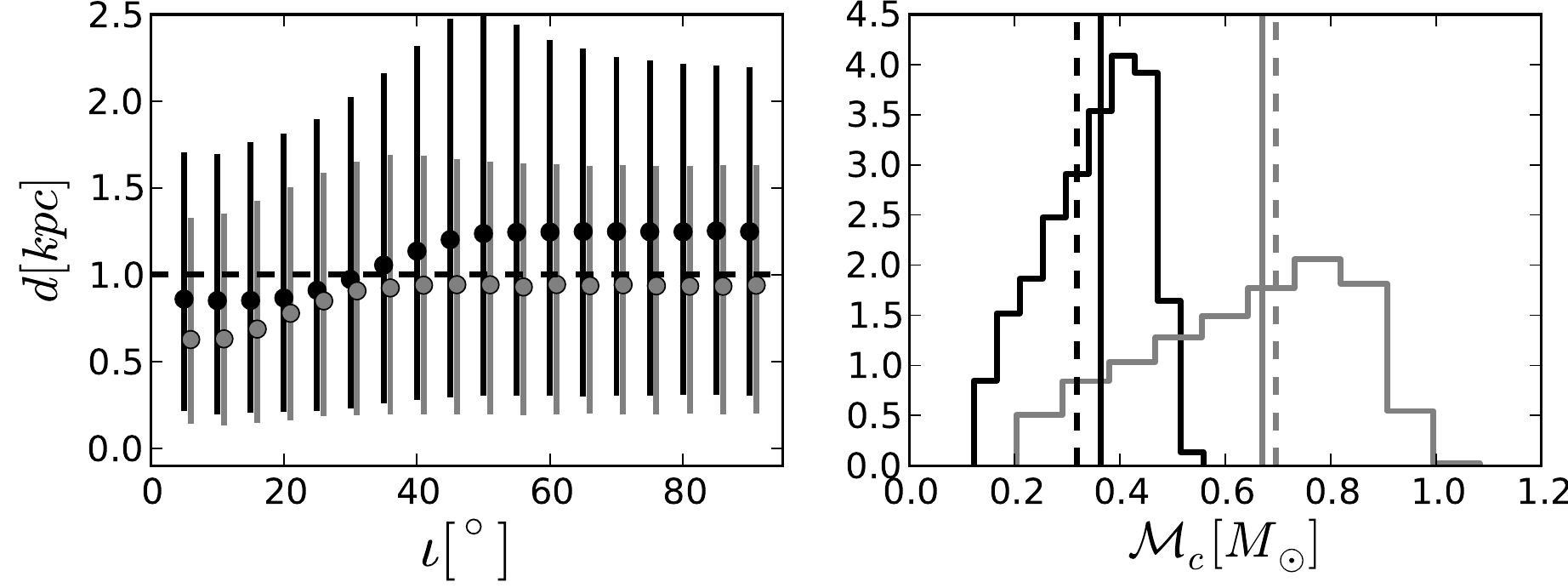}
\caption{\emph{Scenario 2a}: Left: $95$ percentile  in the distance
  given GW data on $\mathcal{A}$ and EM data on $m_1$ for J0651 (in
  black) and the high mass binary (in grey). Note that $\mathcal{M}_c$
  distribution is same at all 
  inclinations for each binary. Right: Comparison of $\mathcal{M}_c$
for J0651 (in black) and for the high mass system (in grey) where
solid lines are the medians of the distributions and dashed lines are
the real values. } 
\label{fig:group2a}
\end{figure*}
\subsection{Scenario 1: EM observation of the distance }
Measuring distances accurately is made feasible by the \textit{Gaia}
mission \citep{2012Ap&SS.341...31D}, a new astrometric
satellite. \textit{Gaia} is expected to measure stellar parallaxes of
millions of stars with $\mu$arcsec accuracy depending on how bright a
star is. For example at 1 kpc, J0651 (g = 19.1 mag) would have a
  $\sim 300\mu$arcsec accuracy in the parallax measurement
  corresponding to a fractional accuracy in distance of $\sim 10\%$
\citep[e.g.][]{2009IAUS..254..475B}. There is also some indication of
the distance of the binary from
its absolute magnitude. The  uncertainties in $d$ from such
measurements are also of the order of several percent or $10\%$ for the
case of J0651\citep{2011ApJ...737L..23B}. 

A sole EM measurement of the distance of a WD binary might be possible
in cases where the system is identified as a WD binary but it is too
faint to measure other parameters. For instance from the wide-field
surveys it is often possible to identify WD from their
colors \citep{2013MNRAS.434.2727V}. Given the distance and the GW uncertainty in amplitude,
we can trivially solve for the chirp mass, $\mathcal{M}_c$ using
Eq.\,\ref{eq:amp}. The resulting probability distribution functions
(pdf) are computed by randomly drawing points from the given
distributions and computing the parameter of interest for each
draw. The $95$ percentiles in the $\mathcal{M}_c$ are shown in the
left panel of Figure~\ref{fig:group1} for J0651 (in black) and the
high mass binary (in grey) as a function of inclination. The dashed
lines (in black for J0651 and in grey for the high mass binary) are
the real values. The decreasing medians of the chirp mass with
inclination follows from the GW distributions in the amplitude that is
shown in Figure~\ref{fig:group0_exp} where the median $\mathcal{A}$ is
overestimated for $\iota = 5^{\circ}$  (in thin black lines) compared to that of $\iota =
90^{\circ}$ (in thick black lines). For a fixed distribution in distance the corresponding
distribution of $\mathcal{M}_c$ is therefore overestimated for $\iota
= 5^{\circ}$ shown in the right panel of Figure~\ref{fig:group1}
compared to that of $\iota = 90^{\circ}$. The $95$ percentiles of the
chirp masses for both J0651 and the high mass binary are affected by
these overestimated medians of the amplitudes at lower inclinations
which cause significant offsets of the $\mathcal{M}_c$ from their
respective real values as can be seen in the left panel of
Figure~\ref{fig:group1}. Thus at lower inclinations where the medians
in the amplitude are overestimates, the $95$ percentiles in the chirp
mass can be interpreted as upper limits of the chirp mass. In order to
calculate reliable constraints in $\mathcal{M}_c$ at these small
inclinations we have to do full (Bayesian) data analyses that takes into
account the physical priors and gives us a better estimate of the expected posterior
distributions in the desired parameters. The $95$ percentile in
$\mathcal{M}_c$ for both systems decrease as a function of inclination
as is expected from the propagation of uncertainty where
$\sigma_{\mathcal{M}_c} \propto \sigma_{\mathcal{A}}$. Thus, knowing
distance from EM observation gives us an estimate of the chirp mass where
the constraints are tighter for the higher inclination (eclipsing)
systems.
\subsection{Scenario 2: EM observations of single-lined
    spectroscopic binary}
Some measurements are unique to EM observations such as the radial
velocity $K_1$, of one of the components ($m_1$) of the binary:
\begin{equation}
K_1 = \sin \iota \: \frac{m_2}{(m_1+m_2)^{2/3}} \:
\: \left(\frac{2\pi \: G}{P_{\mathrm{orb}}}\right)^{1/3}, 
\label{eq:RV}
\end{equation}
 which can be used to measure inclination. We adopt the
 convention from the optical studies of the binary sources where the
 primary mass, $m_1$ is the brighter object and  
 the dimmer secondary mass, $m_2$. Note that the inclination measurement 
 from the GW data analysis, $\iota[\mathrm{GW}]$ and from the radial
 velocity equation above i.e. $\iota[\mathrm{RV}]$ are two 
 independent measurements for the same system. We will show that these
 two are anti-correlated below in Sect. 3.2.3, yielding radial
 velocity measurements very useful. 
\subsubsection{Scenario 2a: EM data on $m_1$}
Before looking at a real single-lined spectral binary we first consider
the case that only the mass $m_1$ is known from the EM data. This is a
viable scenario when determining $K_1$ is impossible and we may get an
estimate of the primary mass from the photometry or the spectra.
Assuming a double WD 
system, we take a uniform distribution for $m_2$, which together with
the given $m_1$ constrains the $d$. The estimates of distance with their
corresponding $95$ percentiles as a function of inclination are shown
in Figure~\ref{fig:group2a} for both the J0651 (in black) and the high
mass binary (in grey). The real value of distance is shown in the dashed
(black) line. The offsets of the medians in the distance at low
inclinations for both the binary systems can be explained in a similar way
as in the previous sections, which is due to the overestimated medians
of $\mathcal{A}$ at lower inclinations as shown in the left panel of
Figure~\ref{fig:group0_exp}. Additionally, the significant discrepancy
between the median distance for J0651 vs. the high mass binary (at
$\iota \geq 40^{\circ}$) is again due to the over-estimated value of the
$\mathcal{M}_c$ for J0651 assuming a uniform distribution $m_2$
distribution. This is shown in the right panel of
Figure~\ref{fig:group2a} where the vertical dashed lines are their
corresponding true values of the $\mathcal{M}_c$ and the vertical
solid lines are the medians of the corresponding distributions. The
simulated distribution of $\mathcal{M}_c$ from an EM measurement of
$m_1$ with a Gaussian width in its uncertainty 
together with an assumed uniform distribution in the unknown $m_2$
results in an overestimated median of the $\mathcal{M}_c$ for J0651
compared to that of the high mass binary. This propagates
in overestimating the median $d$ for J0651 at higher inclinations
unlike for the high mass binary since its median $\mathcal{M}_c$ is
slightly underestimated. The flat priors on $m_2$ is affecting this
and if we already know the secondary mass is low, we may take a
distribution in $m_2$ weighted towards lower masses and that will
affect the constraints obtained in the $d$. The constraints in the
distance from Figure~\ref{fig:group2a} can be compared with those in
Figure~\ref{fig:group0} where there was no EM 
information on any of the masses: the upper limits on $d$ for both
J0651 and the high mass binary are constrained by a up to factor of
$\sim4$ better when $m_1$ is known for both binaries with $10\%$
accuracy.
\begin{figure*}[t]
\centering 
\includegraphics[width=17cm]{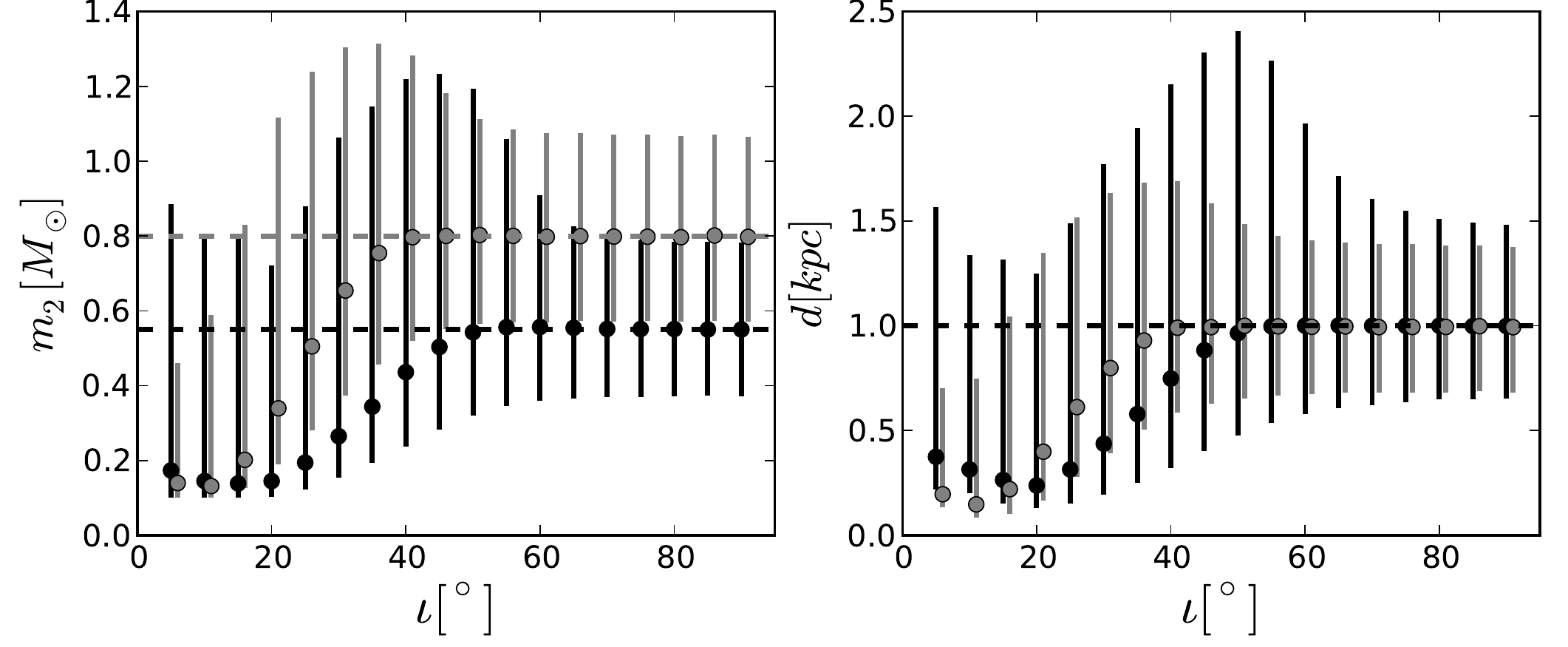}
\caption{\emph{Scenario 2b}: Constraints on the secondary mass and
  distance from combining \emph{single spectroscopic} EM data: $m_1$,
  and $K_1$ with GW data on $\mathcal{A}$, $\iota$ for J0651 (black) and the
  high mass binary (in grey).} 
\label{fig:group2b}
\end{figure*}
\subsubsection{Scenario 2b: EM data on $m_1$ \& $K_1$}
In this case we consider EM measurements of a \emph{single-lined
  spectroscopic} binary where resolving one of the masses of the
binary spectroscopically typically provides measurements on the
primary mass and its radial velocity. We assume an uncertainty in
radial velocity amplitude of $10\%$ corresponding to the typical
accuracy of $10$km/s found in the EM measurements \citep[for
e.g.,][]{2006MNRAS.371.1231R}. Given $m_1$ and $K_1$ from the EM data
and inclination from GW data $\iota[\mathrm{GW}]$, we can numerically
solve for $m_2$ via the $K_1$ formulation in
Eq.\,\ref{eq:RV}. Assuming it is a WD, the $m_2$ is restricted to lie
in $[0.1-1.4] M_{\odot}$. Then a fixed pair of
[$\mathcal{A}$, $\iota[\mathrm{GW}]$] and the masses give us a
distance. We calculate the resulting distributions in $m_2$ and the
distance 
from the Gaussian distributions of $m_1$ and $K_1$ about their typical
EM uncertainties and GW distributions in the inclination and
amplitude. The $95$ percentile in the secondary mass and the distance
are shown in Figure~\ref{fig:group2b} as a function of inclination
for both J0651 (in black) and the high mass binary (in grey). Like
in the scenarios discussed above, for the lowest inclinations, the
over estimated FIM uncertainties of $\mathcal{A}$ propagates into
erroneous constraints on $m_2$, and $d$. Thus, at lower inclinations we have to use
Bayesian methods to get their accurate GW uncertainties. Observe that
the $95$ percentile in the $m_2$ and the distance roughly similar and
large from $5^{\circ}<\iota<45^{\circ}$. This is again due to the
influence of the GW distributions in $\mathcal{A}$ at the lower
inclinations, which have uniform distributions resulting into
over-estimated medians (see Figure~\ref{fig:dist}). However, for
$\iota>45^{\circ}$ the uncertainties for both $m_2$, and $d$ decrease
with inclination and their medians stabilize at the true values. This
is caused by the fact that at higher inclinations, the medians of GW
amplitudes are close to the true values of the systems where the
constraints on the GW parameters are also tighter with increasing
inclination. Thus, the decreasing uncertainties in
$\iota[\mathrm{GW}]$ as a function of $\iota$ (see right panel of
Figure~\ref{fig:uncertainties}) should result in the same behavior of
$\sigma_{m_2}$ via Eq.\,\ref{eq:RV}. Since distance is computed using
these $m_2$, the same behavior holds for the distance in the right panel. 
\begin{figure*}
\centering
\includegraphics[width=17cm]{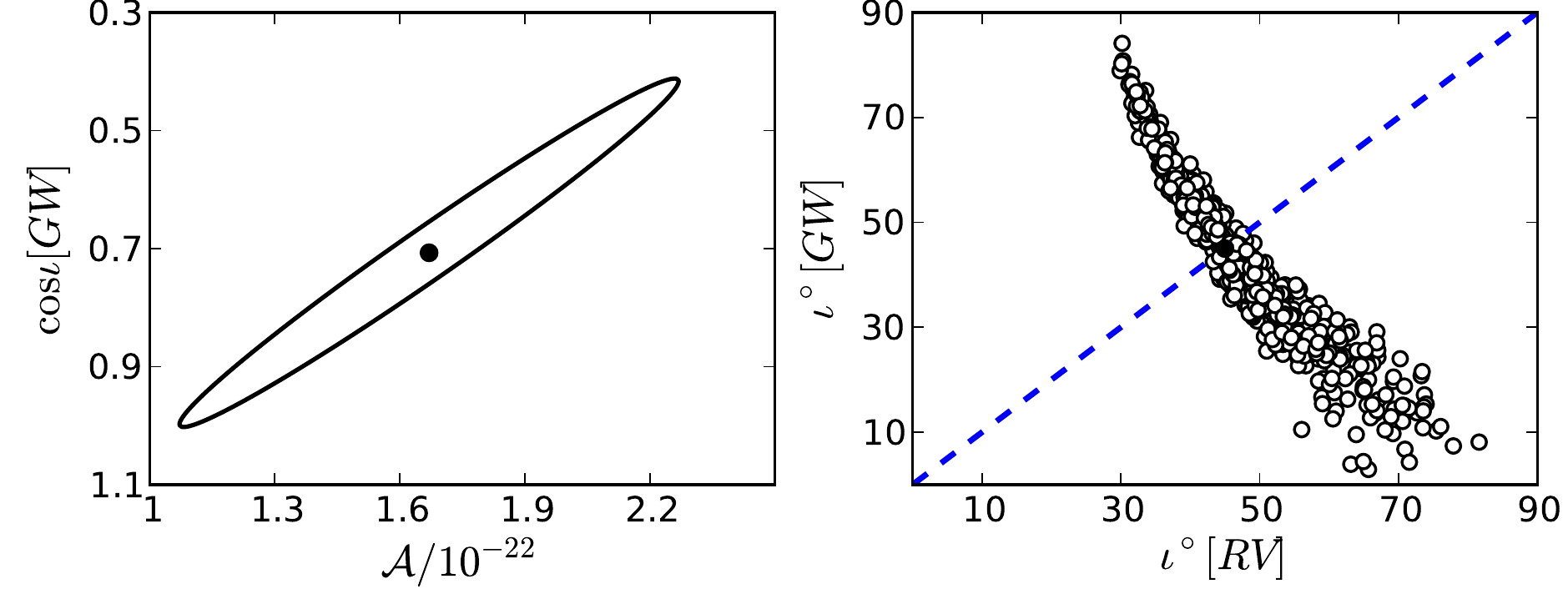}
\caption{Relation between inclination from GW observation
  vs. that from EM observations. \emph{Left} 2-d error ellipse from the GW data analysis in
  amplitude and $\cos\iota$ for J0651 with $\iota = 45^{\circ}$.
\emph{Right}: Relation between inclination from
  the left panel and inclination from Eq.\,\ref{eq:RV}.}
\label{fig:rv_gw_relation}
\end{figure*}
\begin{figure*}
\centering 
\includegraphics[width=17cm]{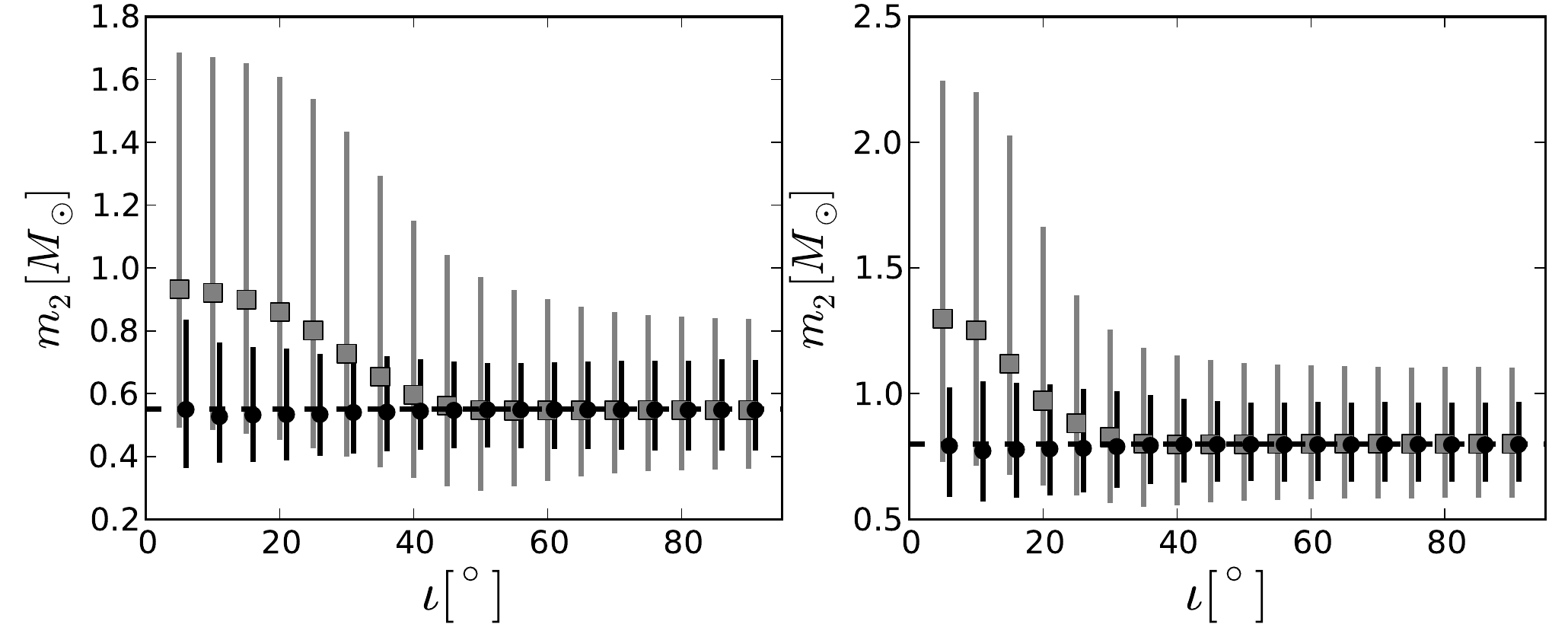}
\caption{\emph{Scenario 2c}: Same scenario as in
  Figure~\ref{fig:group2b} with an additional EM measurement on
  the distance. Left: $2-\sigma$ uncertainties for the secondary mass
  for J0651 where the grey colored lines are constraints from EM information on
  $m_1$, $d$ and GW $\mathcal{A}$. These reduce to the tighter 
  constraints shown in black lines when EM data on $K_1$ is also used
  (see text). The dashed line in red is the real value of $m_2$.
  Right: Same for the high mass binary. } 
\label{fig:group2c}
\end{figure*}
\subsubsection{Scenario 2c: EM data on $m_1$, $K_1$ \& $d$}
Here the EM measurements of a \emph{single-lined spectroscopic} binary
in the previous subsection is complemented with a distance measurement
from \emph{Gaia} or from an estimate of the absolute magnitude. From
the primary mass $m_1$, distance and the amplitude we 
immediately get the secondary mass, $m_2$. We will call this as the
preliminary $m_2$ since this can be further improved by folding in the
radial velocity measurement. As mentioned before the radial
  velocity measurement essentially provides an independent measurement
  of the inclination via Eq.~\ref{eq:RV}. This can be seen in the following way:
The GW parameters of the non-eclipsing
J0651 are: $\mathcal{A}_0,$ $\iota _0$ $= 1.67\times 10^{-22},
45^{\circ}$ whose VCM uncertainties are:
$\sigma_{\mathcal{A}}/\mathcal{A} = 0.231$, and $\sigma_{\iota} =
0.75$ rad respectively. We also
take a fixed radial velocity, $K_0$ corresponding to $m_1$, $m_2$
(listed in Table 1), and $\iota _0$. The 2-d Gaussian distribution
from GW data with $1-\sigma$ uncertainties for these parameters is
shown in the left panel of Figure~\ref{fig:rv_gw_relation}. For each
randomly selected pair of $[\mathcal{A}, \cos\iota[\mathrm{GW}]]$ and for
a fixed $m_1$, and $d$, we can solve for the $m_2$ from
Eq.\,\ref{eq:amp}. Using this $m_2$ for that fixed $m_1$ and $K_0$, we
solve for $\iota[\mathrm{RV}]$. For many points randomly picked in the
$[\mathcal{A}, \cos\iota[\mathrm{GW}]]$ space the computed
$\iota[\mathrm{RV}]$ are compared with the corresponding
$\iota[\mathrm{GW}]$ in the right panel. The inclinations
measured in two ways roughly anti-correlate. However we know that values of
$\iota[\mathrm{RV}]$ that are different from $\iota[\mathrm{GW}]$
cannot be true. Thus, constraining the inclination of the system in a
small area around $45^{\circ}$ along the diagonal line in the right
panel also constrains 
$m_2$ and the amplitude. We make use of this in the case
considered in this subsection.
The preliminary $m_2$ and their $95$ percentiles computed from
EM data on $m_1$, $d$, and the GW data on $\mathcal{A}$ as a function
of inclination is shown in Figure~\ref{fig:group2c} in grey
lines in the left panel for J0651. The same for the high mass binary
is also shown in the right panel in grey lines. From this $m_2$, given
$m_1$, and $\iota[\mathrm{GW}]$, the radial velocity, $K_{\mathrm{GW}}$
is computed which is compared with the $K_1$ from the EM data. Since the
EM measured $K_1$ is more precise, we keep 
the subset of those $K_{\mathrm{GW}}$ and the respective
$\iota[\mathrm{GW}]$ weighted with a probability distribution function of the
$K_1$ given by:
\begin{equation}
\mathcal{P}_i = \frac{1}{\sqrt{2\pi\: \sigma_{K_1}}} \: \exp
\left(\frac{-0.5 \:  (K_{1,i}[\mathrm{GW}] - K_1)^2}{\sigma_{K_1}} \right) \: dK_1.
\end{equation}
The final reduced $95$ percentiles in $m_2$ are shown in black lines
for J0651 in left panel, and the same is shown for the high mass binary in the right
panel. Observe that the uncertainties in $m_2$ calculated in this way
for lower inclinations is the similar to those at the higher
inclinations. Thus, the advantage of folding in 
$K_1$ measurement is \emph{especially} useful for lower inclination systems
with S/N $\sim10$, where large GW uncertainties
$\mathcal{A}$ influence the constraints of the physical parameters in
question. Furthermore, the constraints in $m_2$ can also be
compared with the previous case in Figure~\ref{fig:group2b} where we
find that for the single-lined spectroscopic binary, knowing its distance to
$10\%$ significantly improves knowledge of the secondary mass at lower
inclinations.

The key point in Scenarios 2b and 2c is that not all the $[\mathcal{A},
\iota[\mathrm{GW}]]$ pairs are consistent with the EM
observations. Therefore both constraints on the GW data and other
parameters also constrain the GW error ellipses. The $2-\sigma$
uncertainties in the GW amplitude and GW inclination for these
scenarios are shown in the Appendix. 
\begin{figure*}[t]
\centering 
\includegraphics[width=17cm]{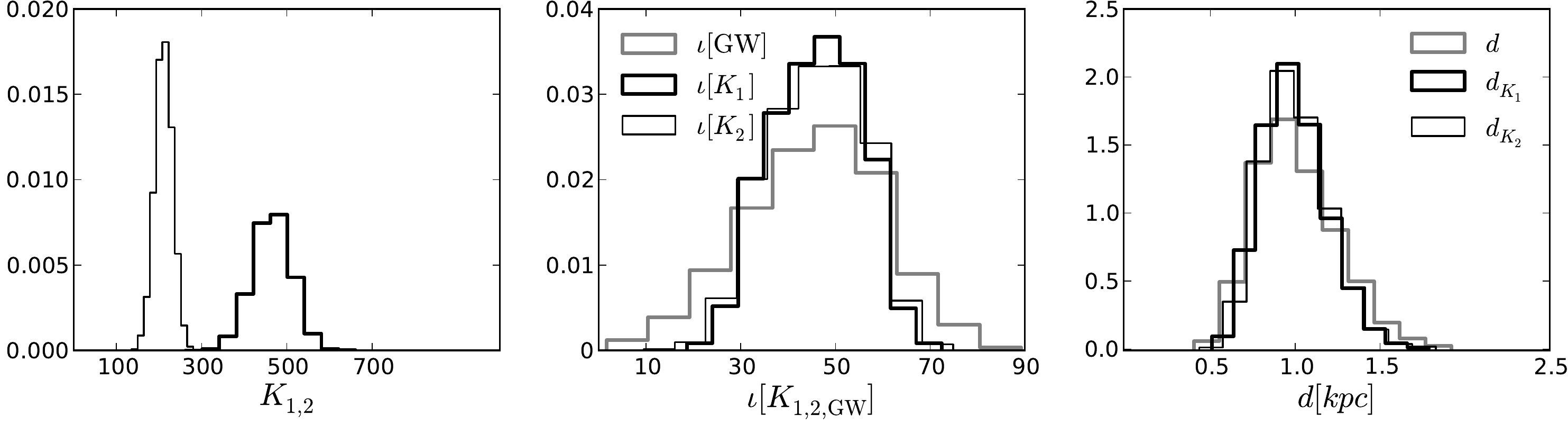}
\caption{\emph{Scenario 3}: Left: Distributions of the radial
  velocities from EM data for J0651 masses with the binary orientated
  at $\iota =45^{\circ}$. Middle: Given EM data on $m_1,
  m_2$, and the corresponding $K_1, K_2$, the inclinations are
  calculate using Eq.\,\ref{eq:RV} which is compared with inclination
  from GW data. Right: Constraints on the distance obtained solving
  Eq.\,\ref{eq:amp} with: EM data on $m_1$, $K_1$, $m_2$, $K_2$ and 
  GW data on $\mathcal{A}$, $\iota$. }
\label{fig:group3_exp}
\end{figure*}
\begin{figure*}
\centering
\includegraphics[width=17cm]{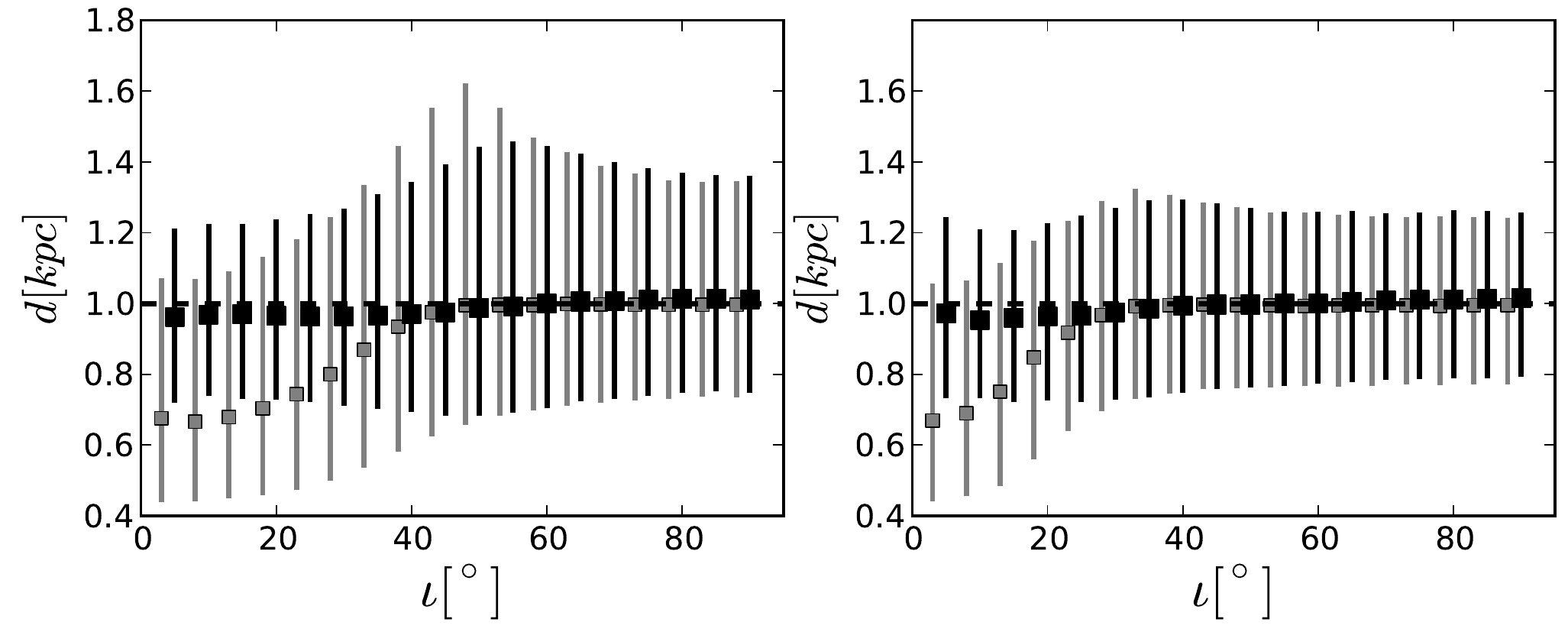}
\caption{\emph{Scenario 3}: Same as in Figure~\ref{fig:group3_exp} but
  for all inclinations for J0651 in left panel and the for the 
  high mass binary in right panel. The constraints in black
  lines are from using $\iota[K_1]$ and the constraints in grey are
  from using $\iota[\mathrm{GW}]$. The dashed line (in black) is the
  real values of the $d$.}
\label{fig:group3}
\end{figure*}
\subsection{Scenario 3: EM data on $m_1$, $K_1$, $m_2$ \& $K_2$}
In this section we consider EM observations of a \emph{double-lined 
spectroscopic} binary which translates to a set of measurements in the mass
and radial velocity for each of the components: $m_1$, $K_1$, and
$m_2$, $K_2$. Given the two masses and GW measurement on the amplitude
we can immediately compute a preliminary distance. Additionally, we
can also derive two sets of inclinations independently from the individual
radial velocities and the masses, $\iota_{\mathrm{K_1}},
\iota_{\mathrm{K_2}}$ from Eq.\,\ref{eq:RV}. These inclinations can
be compared with the one measured from GW data,
${\iota}[\mathrm{GW}]$. At lower inclinations, large
uncertainties in ${\iota}[\mathrm{GW}]$ essentially imply that those
systems' inclinations are undetermined and this also affects the
amplitude due to the strong correlation between them. Thus, the
independent estimates of $\iota_{\mathrm{K_1}}, \iota_{\mathrm{K_2}}$
from the EM data can be useful in constraining the GW amplitude. This
reduced amplitude will further constrain the distance which is
shown in the third panel in Figure~\ref{fig:group3_exp}. In the figure both
the observed $K_1, K_2$ are shown in the left panel in thick and thin black lines
respectively. The inclination and the distance given the GW amplitude
and both the masses are shown in grey line in 
the middle and right panels respectively. Both the inclination and distance
derived from $K_1$ and $K_2$ are plotted in thick and thin black lines
respectively. Observe that a $10\%$ fractional error in each $K_1$ and $K_2$ translate
into similar uncertainties of the distance and thus in the following
figures we show the constraints
from using $K_1$ data only. The constrained distances estimated in this 
way as a function of inclination is shown in Figure~\ref{fig:group3}
for J0651 in left panel (also in black) and for the high mass binary in the right
panel (in black). The grey lines in both the panels are the
$95$ percentiles in $d$ using only the masses from the EM data and the GW
amplitude. Observe that at lower inclinations knowing masses and a radial
velocity can improve the constraint in distances significantly. The
uncertainties are smaller for J0651 at lowest inclinations because the
relative $10\%$ uncertainties in the $K_1$ have lower absolute
uncertainties that propagate into the uncertainties of the distance. 

Note that typically in practice EM data provides measurements of both the
masses and only one of the radial velocities with $\sim 10\%$
precision. From the radial velocity formulation we have the relation:
relation $m_1/m_2 = K_2/K_1$ which can be used to compute the
remaining $K_i$. This provides a consistency check between EM and GW
data. The EM data can be used to derive inclination measured from the
radial velocities, $\iota[\mathrm{RV}]$ which can be verified against
the $\iota[\mathrm{GW}]$ as shown in the middle panel in
Figure~\ref{fig:group3_exp}.
\begin{figure*}
\centering 
\includegraphics[width=17cm]{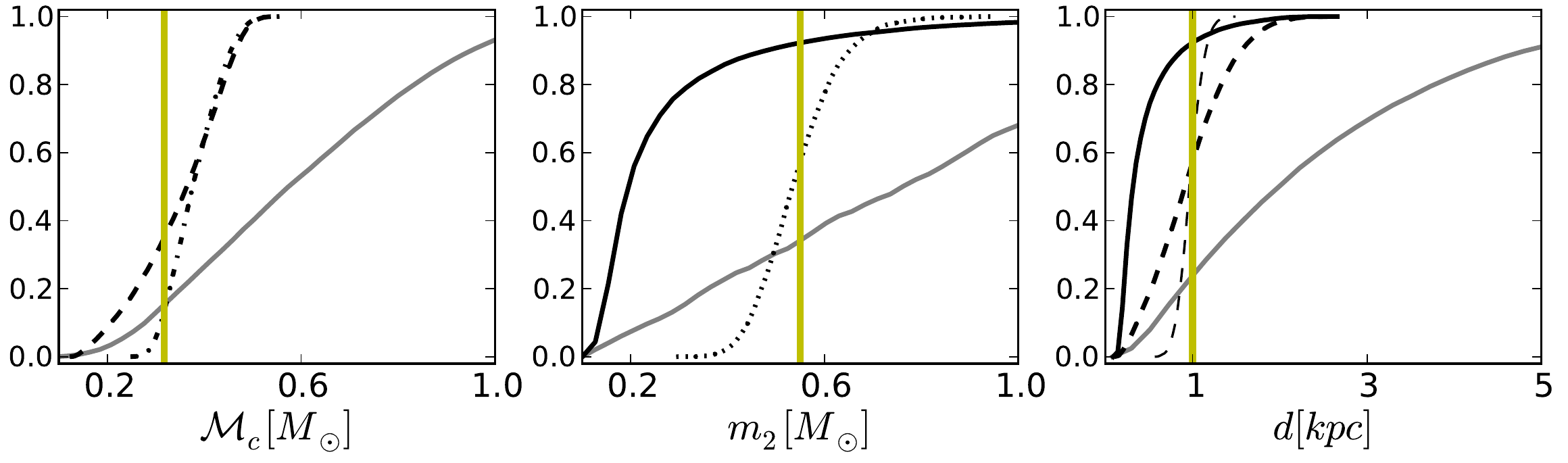}
\caption{Comparison of normalized CDFs in $\mathcal{M}_c$, $m_2$, and $d$
  for all the scenarios above for J0651 with $\iota =
  25^{\circ}$. The vertical lines (in yellow) in all the panels are
  the true values of the parameters. The solid curves in grey are CDFs
for the parameters when only GW data is available. Curves in
dash-dotted lines are constraints for \emph{Scenario 1 (known
  distance $d$)}, dashed curves
are for \emph{Scenario 2a (known
  primary mass $m_1$)}, solid curves are for \emph{Scenario 2b (known
  primary mass $m_1$ and radial velocity $K_1$)},
dotted curves are for \emph{Scenario 2c (known
  $m_1$, $K_1$ and $d$)} and thin-dashed lines are for
\emph{Scenario 3 (known
  both masses $m_1, m_1$ and radial velocities $K_1, K_2$)}.}
\label{fig:compare_25}
\end{figure*}
\begin{figure*}
\centering 
\includegraphics[width=17cm]{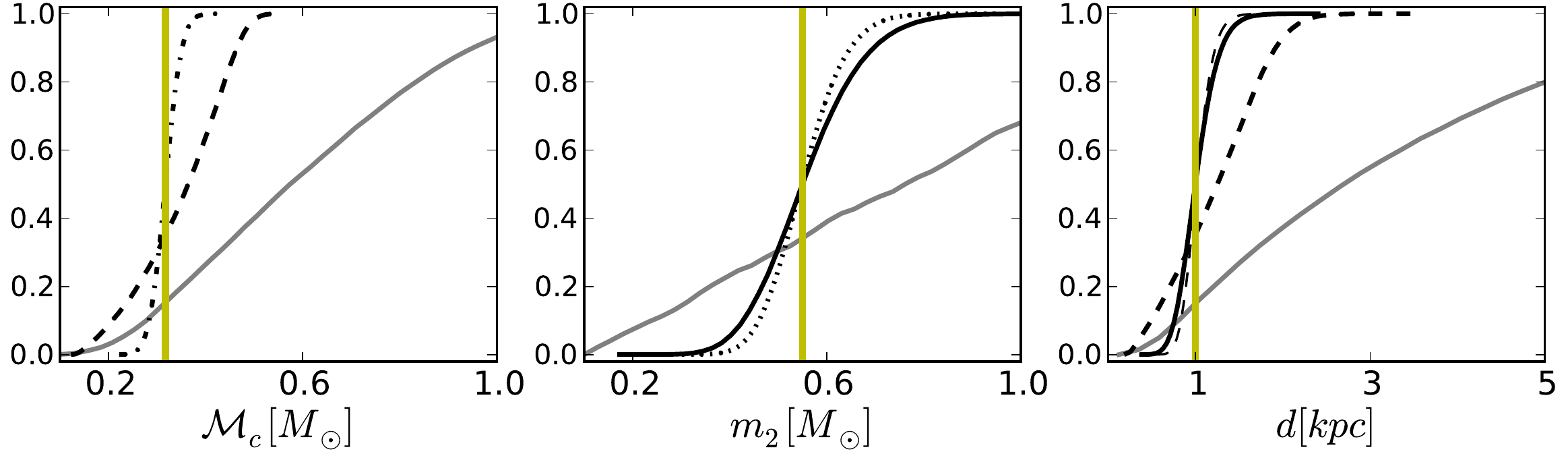}
\caption{Same as in Figure~\ref{fig:compare_25}, for $\iota =
  85^{\circ}$.}
\label{fig:compare_85}
\end{figure*}
\section{Conclusions}
We have quantified the possible constraints/improvements in the
physical parameters of the white-dwarf (WD) binaries that are observable
by the eLISA detector in the future when combined with the EM data. We do
this for the binary parameters that are astrophysically interesting
(masses and distance). For the GW observations from eLISA,  
we calculate the source's variance-covariance matrix using the Fisher
methods where the Galactic binary source is described by seven
parameters (or eight if $\dot{f}$ is measurable). We have taken J0651
and a higher mass binary in our analyses where J0651 is a
\emph{verification} source for eLISA. We consider various possible
cases depending on the availability of the EM measurements and combine
those with GW uncertainties in the amplitude and inclination in order
to solve for the unknown parameters as a function of inclinations for
both J0651 and the high mass binary. For clarity we list all the cases
 below:
\begin{enumerate}
\item \emph{GW data only}: Assuming a double white-dwarf system this scenario
  somewhat constrains the distance.
\item \emph{Scenario 1}: GW data + distance $d$: This scenario constrains the 
  chirp mass $\mathcal{M}_c$.
\item \emph{Scenario 2a}: GW data + primary mass $m_1$: This scenario constrains the 
  chirp mass and the distance.
\item \emph{Scenario 2b}: GW data + single-lined spectroscopic binary
  i.e. $m_1$, $K_1$: This scenario constrains the 
  secondary mass $m_2$ and the distance.
\item \emph{Scenario 2c}: GW data + single-lined spectroscopic binary + $d$:
  This scenario also constrains the secondary mass $m_2$. 
\item \emph{Scenario 3}: GW data + double-lined spectroscopic binary
  i.e. $m_1$, $K_1$ and $m_2$, $K_2$:
  This scenario constrains the distance. 
\end{enumerate}
 All the $1-\sigma$ EM accuracies are taken to be
$10\%$ of the real/measured values which is inspired by several EM
observations. We compare below the
constraints in the physical parameters of interest: secondary mass
$m_2$, chirp mass $\mathcal{M}_c$ and the distance $d$ as a function
of the scenarios depending on the EM information available. Since the GW
parameter uncertainties are significantly different for a low inclination
(face-on) orientation than for a high inclination (edge-on)
orientation, we do the comparison for a non-eclipsing J0651 with
$\iota =25^{\circ}$ and an almost eclipsing J0651 with $\iota =85^{\circ}$  in
Figures~\ref{fig:compare_25} and ~\ref{fig:compare_85} respectively
and conclude the following:
\begin{enumerate}
\item \emph{Constraints on chirp mass, $\mathcal{M}_c$:} In the left panels of
  Figures~\ref{fig:compare_25} and~\ref{fig:compare_85}, EM data on $d$
  constrains the 95 percentile of the system's chirp mass (dash-dotted line) 
  to $0.38^{+0.11}_{-0.09} M_{\odot}$ and $0.32^{+0.05}_{-0.05} M_{\odot}$ for face-on and eclipsing
  J0651 respectively. EM data on $m_1$ constraints the $\mathcal{M}_c$ 
  (in thick-dashed line) to $0.36^{+0.13}_{-0.21} M_{\odot}$ which does not depend on the
  inclination. The normalized cumulative distributions (CDF) of the constraints on the
  distance are compared to that from GW data
  only which is shown in the grey line in both panels.
\item \emph{Constraints on secondary mass, $m_2$:} In the middle panels of
  Figures~\ref{fig:compare_25} and~\ref{fig:compare_85}, EM data on
  the $m_1, K_1$ constrain the 95 percentile of secondary mass, $m_2$
  to $0.19^{+0.69}_{-0.07} M_{\odot}$ and 
  $0.55^{+0.23}_{-0.18} M_{\odot}$ for face-on and 
  eclipsing J0651 respectively (shown in solid lines). The same set of
  data complemented with the distance further constrain the 95
  percentile in $m_2$ with 
  $0.55^{+0.18}_{-0.13} M_{\odot}$ and $0.55^{+0.16}_{-0.13} M_{\odot}$ for face-on and
  eclipsing J0651 respectively (shown in dotted lines). For
  comparison, the CDF of $m_2$ using only the GW data is shown in grey. 
\item  \emph{Constraints on distance, d:} In the right panels, of
  Figure~\ref{fig:compare_25} and~\ref{fig:compare_85}, EM data on $m_1$
  constrains the distance to $0.91^{+0.98}_{-0.69}$ kpc and $1.25^{+0.95}_{-0.95}$ kpc for
  face-on and eclipsing J0651 respectively (in thick-dashed lines). EM
  data on the $m_1, K_1$ constrain the 95 percentile in $d$ with 
  $0.32^{+1.17}_{-0.16}$ kpc and with $0.99^{+0.49}_{-0.35}$ kpc accuracy for
  face-on and eclipsing J0651 respectively (in solid lines). EM
  data on $m_1, m_2, K_1$ and $K_2$ constrain the 95 percentile in
  $d$ to $0.96^{+0.29}_{-0.24}$ kpc and $1.01^{+0.35}_{-0.26}$ kpc for face-on and eclipsing
  J0651 respectively (in thin-dashed line). For comparison, the CDF
  of $d$ using only the GW data and the assumption that the masses are
  WDs is shown in grey.
\end{enumerate}
Thus, knowing distance and/or radial velocity of the primary component
can significantly improve our knowledge of the binary system. These
constraints change as a function of inclination of the 
binary that is shown in previous sections. In a forthcoming paper we will
address the effect on these improvements by including the (possible)
EM measurement of rate of change of the orbital period.
\begin{acknowledgements}
This work was supported by funding from FOM. We are very grateful to
Michele Vallisneri for providing support with the \textit{Synthetic LISA} and
\textit{Lisasolve} softwares. 
\end{acknowledgements}
\bibliographystyle{apj}
\bibliography{/Users/swetashah/Documents/writings/literature_binary_science,/Users/swetashah/Documents/writings/literature_data_analysis}
\appendix
\section{Variance-covariance matrixes of J0651}
We have listed the VCM matrices for the J0651 system with eclipsing
and non-eclipsing configurations in
our analysis. There are 7 parameters that describe them which are
listen in the first row of the matrices below and for each binary, the
values are listed in the row with $\theta_{i}$. The diagonal elements
are the absolute uncertainties in each the 7 parameters and the
off-diagonal elements are the normalized correlations, i.e.  
$\mathrm{c}_{ii} = \sqrt{\mathcal{C}_{ii}} \equiv \sigma_{i},$
$\mathrm{c}_{ij}
=\frac{\mathcal{C}_{ij}}{{\sqrt{\mathcal{C}_{ii}\mathcal{C}_{jj}}}}.$
The strong correlations between parameters (i.e. whose magnitudes are
$\ge 0.9$) are marked in bold in the VCMs below. These correlations
have been explained in Paper I. \\ 
VCM 1: Eclipsing J0651 ($\iota = 5^{\circ}$), S/N $= 10.5$.  \\\\
\resizebox{\linewidth}{!}
{
\bordermatrix{~ & \mathcal{A} & \phi_0 & \cos\iota & f & \psi & \sin\beta & \lambda \cr
\theta_i&1.67\times10^{-22} & \pi & 0.007& 2.61\times10^{-3}& \pi/2&-0.08 &2.10\cr \hline \cr
\mathcal{A}  &0.08\times10^{-23}& -0.0& \;\;\;0.0& \;\;\;0.01& \;\;\;0.02& \;\;\;0.03& -0.06\cr
\phi_0& & \;\;\;0.907&-0.01& \mathbf{-0.91}&  \;\;\;0.01& \;\;\;0.11& \;\;\;0.08\cr
\cos\iota && &\;\;\;0.172&\;\;\;0.01& -0.01& \;\;\;0.07& -0.33\cr
f & &&& 2.982\times10^{-10}& -0.01& -0.08& -0.15\cr
\psi & &&&&\;\;\;0.035& -0.02& \;\;\;0.05\cr
\sin\beta &&&&& & \;\;\;0.059& \;\;\;0.08\cr
\lambda&&&&&&&  \;\;\;0.017 \cr  \cr}
}\\
VCM 2: Not-eclipsing J0651 ($\iota = 45^{\circ}$), S/N $= 24$.\\
\resizebox{\linewidth}{!}
{
\bordermatrix{~ & \mathcal{A} & \phi_0 & \cos\iota & f & \psi & \sin\beta & \lambda \cr
\theta_i&1.67\times10^{-23}& \pi & 0.707& 2.61\times10^{-3}&\pi/2&-0.08 &2.10\cr \hline \cr
\mathcal{A}  &3.86\times10^{-23}& \;\;\;0.03& \mathbf{-0.98}& -0.02& \;\;\;0.03& -0.13& \;\;\;0.35\cr
\phi_0& & \;\;\;0.739&-0.03& -0.19& \;\;\;0.16& \;\;\;0.15& \;\;\;0.10\cr
\cos\iota && &\;\;\;0.19&\;\;\;0.02& -0.01& \;\;\;0.13& -0.36\cr
f & &&& 1.688\times10^{-9}& \mathbf{-0.98}& -0.06& -0.21\cr
\psi & &&&&\;\;\;0.36&\;\;\;0.13& 0.07\cr
\sin\beta &&&&& & \;\;\;0.031& -0.13\cr
\lambda&&&&&&&  \;\;\;0.009 \cr  \cr}
} 
\section{Constraints in $\mathcal{A}$ and $\iota$ of J0651}
\begin{figure*}[!h]
\centering 
\includegraphics[width=\columnwidth]{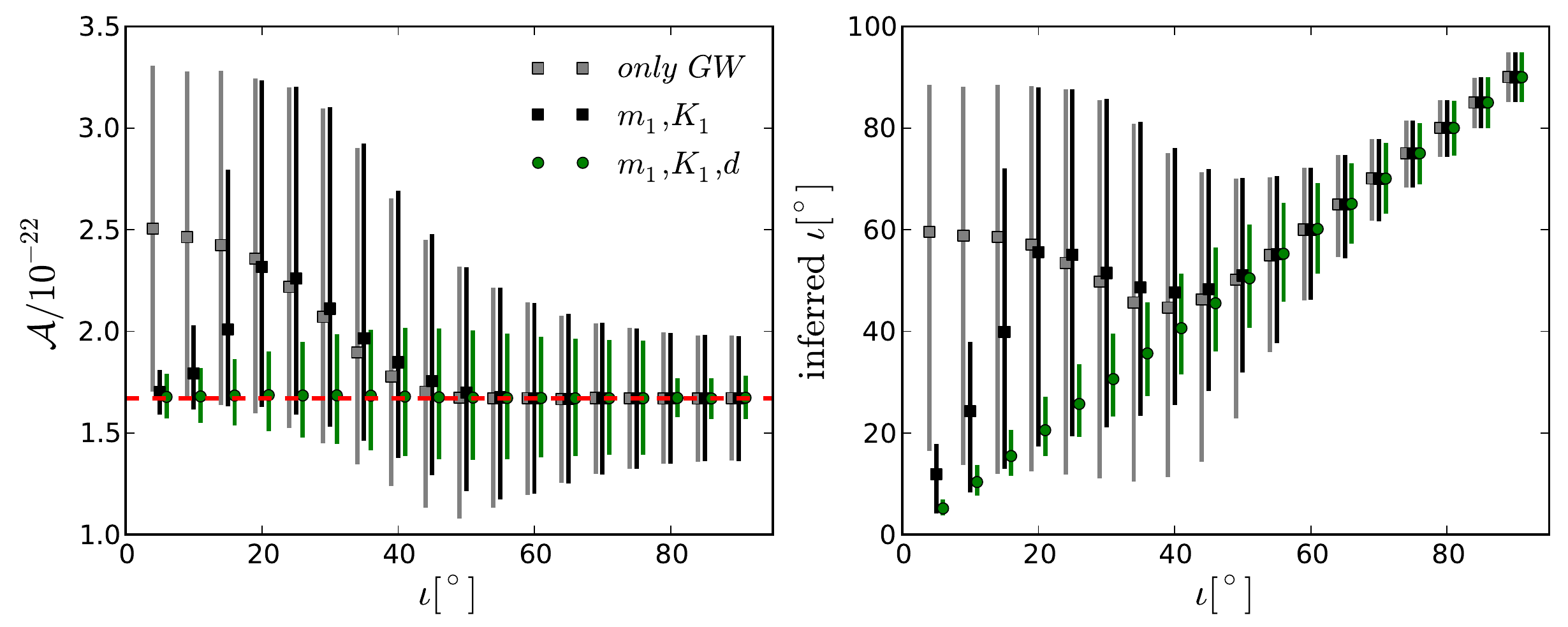}
\caption{$95$ percentiles in the GW amplitude as a function of
  inclination for various sets of EM information as labeled. The real
  value for the amplitude is shown in black dashed-line. }
\label{fig:amp_constraints}
\end{figure*}

Figure~\ref{fig:amp_constraints} shows how the error ellipses of amplitude and
inclination from GW observations reduce using EM observations for the
different scenarios that we have described in Sects. 1 and 2. Knowing
one of the masses (Scenario 
2a) from the EM does not constrain the $[\mathcal{A}, \iota]$ any more than
the GW data alone. In other words the $m_2$ and $d$ are free
parameters to satisfy the amplitude. The $95$ percentiles in the
amplitude are shown in grey in the figure which are the same as the
case where we have GW data only. In fact these constraints in the
amplitude decrease as a function of inclination as expected from the
GW measurements (see Figure~\ref{fig:uncertainties}). Adding an EM
measurement of the measured mass's radial velocity (Scenario 2b) can
constrain the $[\mathcal{A}, \iota]$ slightly or significantly depending on
inclination of the system which are shown in thick black
lines. Finally complementing the mass and radial 
velocity of the brighter companion with the distance to the binary
(Scenario 2c) significantly constraints the $[\mathcal{A}, \iota]$ which is
strongest for the lower inclinations as shown in the figure in thin
black lines. Observe that EM information provide strongest improvements for
low inclination systems where GW uncertainties in the amplitude and
the inclination are very large. 
\section{The distribution of $\mathcal{A}$ and $\iota$ at lower
  inclinations}
\begin{figure*}[!h]
\centering 
\includegraphics[width=\columnwidth]{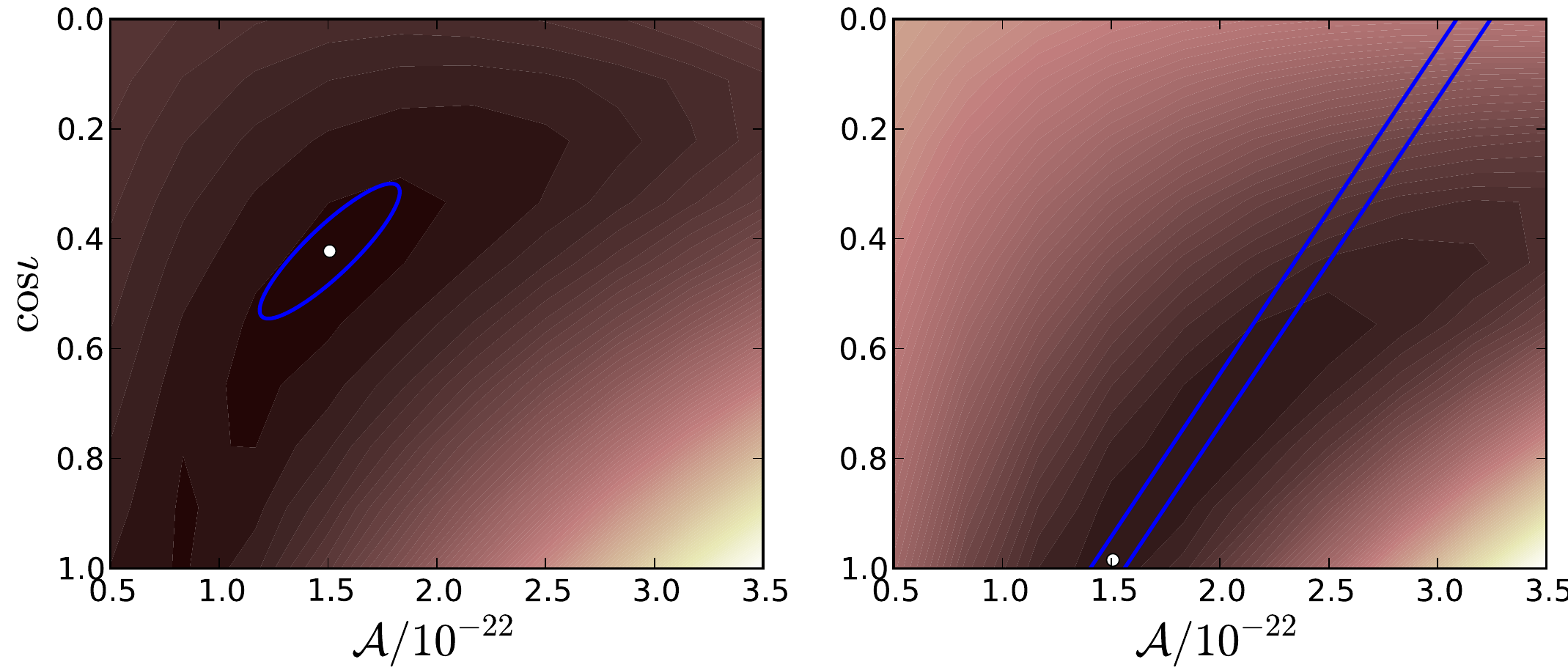}
\caption{Filled contour plot of the 2D $\chi^2$ averaged over 10 noise realisations for an evenly
  distributed grid of $\mathcal{A}$ and $\cos\iota$ compared with the
  $1-\sigma$ error ellipse (shown in thick solid line) from Fisher matrix $\iota=65^{\circ}$ in
  the left-panel and $\iota= 10^{\circ}$ in the right-panel. The
  $\chi^2$ values are represented by the darker to lighter colors
  corresponding to lower to higher values in $\chi^2$.}
\label{fig:chi2}
\end{figure*}

Here we show that while Fisher method gives an estimate on the
parameter uncertainties and correlation between them without
following the posterior in detail, it gives a reasonable estimate of
the above quantities as long as the priors in the parameters are
rectangular (i.e. not Gaussian) and are large enough to preserve the
overall orientation of the posterior. We compute an estimate
of the likelihood with 
a simple $\chi^2$ procedure on a 2D parameter distribution of
$\mathcal{A}_i$ and $\cos\iota_j$, where the $\chi^2 = (1/(N-1))
\Sigma_{t=0,N}(h_0(t)-(h[i,j](t)+n(t))^2$, $h_0$ = true signal,
$h[i,j]$ = signal at a grid point and $n$ is a noise realisation,
$N=$total time samples. For an evenly placed parameters in a $10
\times 10$ grid, we take the average $\chi^2$ computed for 10
different noise realisations. Figure~\ref{fig:chi2} shows the colored
contours of 2D $\chi^2$ distribution for the case of
$\iota=65^{\circ}$ (in the left-panel) where the Fisher uncertainties
are well within the physically allowed bounds. The over-plotted
contour in thick solid line is $1-\sigma$ uncertainty ellipse computed
from Fisher matrix about the true values of $\mathcal{A}$ and
$\cos\iota$ labelled with the white circle. This just shows that the
$\chi^2$ distribution 
follows the shape and the slope of the Fisher distribution roughly,
but not exactly as expected. The same is shown for $\iota=10^{\circ}$
in the right-panel where the uncertainties hit the physical bounds and
both the methods show a sharp cut-off at $\cos\iota=1$. Here we see
that again the Fisher uncertainties and correlation roughly follow 
that of the $\chi^2$, but with truncations at the boundaries. 
The deviation in the top-right is discussed in
\cite{2012A&A...544A.153S}. It was argued that although the results of
Fisher-based uncertainties imply that the $\iota= 5^{\circ}$ system is very similar
to $\iota= 90^{\circ}$, this is unlikely because of the anti-correlation between
$\mathcal{A}$ and $\cos\iota$ at high inclinations. At $\iota \gtrsim
45^{\circ}$ correlation between $\mathcal{A}$ and $\cos\iota$
decreases $\iota$ and the high accuracy in the inclination itself actually
suffices to distinguish the higher inclination systems. Thus we expect
that that $\chi^2$ deviates from the Fisher estimate towards the
top-right region of the Figure~\ref{fig:chi2} in the right-panel. 

\end{document}